\documentclass[twoside,twocolumn,9pt]{article}
\usepackage{extsizes}
\usepackage[super,sort&compress,comma]{natbib} 
\usepackage[version=3]{mhchem}
\usepackage[left=1.5cm, right=1.5cm, top=1.785cm, bottom=2.0cm]{geometry}
\usepackage{balance}
\usepackage{mathptmx}
\usepackage{sectsty}
\usepackage{graphicx} 
\usepackage{lastpage}
\usepackage{amssymb}
\usepackage[format=plain,justification=justified,singlelinecheck=false,font={stretch=1.125,small,sf},labelfont=bf,labelsep=space]{caption}
\usepackage{float}
\usepackage{fancyhdr}
\usepackage{fnpos}
\usepackage[english]{babel}
\addto{\captionsenglish}{%
  
}
\usepackage{array}
\usepackage{droidsans}
\usepackage{charter}
\usepackage[T1]{fontenc}
\usepackage[usenames,dvipsnames]{xcolor}
\usepackage{setspace}
\usepackage[compact]{titlesec}
\usepackage{hyperref}
\usepackage{lipsum}

\usepackage{epstopdf}

\definecolor{cream}{RGB}{222,217,201}

\begin{document}

\pagestyle{fancy}
\thispagestyle{plain}
\fancypagestyle{plain}{
\renewcommand{\headrulewidth}{0pt}
}

\makeFNbottom
\makeatletter
\renewcommand\LARGE{\@setfontsize\LARGE{15pt}{17}}
\renewcommand\Large{\@setfontsize\Large{12pt}{14}}
\renewcommand\large{\@setfontsize\large{10pt}{12}}
\renewcommand\footnotesize{\@setfontsize\footnotesize{7pt}{10}}
\makeatother

\renewcommand{\thefootnote}{\fnsymbol{footnote}}
\renewcommand\footnoterule{\vspace*{1pt}%
\color{cream}\hrule width 3.5in height 0.4pt \color{black}\vspace*{5pt}} 
\setcounter{secnumdepth}{5}

\makeatletter 
\renewcommand\@biblabel[1]{#1}            
\renewcommand\@makefntext[1]%
{\noindent\makebox[0pt][r]{\@thefnmark\,}#1}
\makeatother 
\renewcommand{\figurename}{\small{Fig.}~}
\sectionfont{\sffamily\Large}
\subsectionfont{\normalsize}
\subsubsectionfont{\bf}
\setstretch{1.125} 
\setlength{\skip\footins}{0.8cm}
\setlength{\footnotesep}{0.25cm}
\setlength{\jot}{10pt}
\titlespacing*{\section}{0pt}{4pt}{4pt}
\titlespacing*{\subsection}{0pt}{15pt}{1pt}

\fancyfoot{}
\fancyfoot[LO,RE]{\vspace{-7.1pt}\includegraphics[height=9pt]{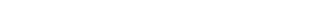}}
\fancyfoot[CO]{\vspace{-7.1pt}\hspace{13.2cm}\includegraphics{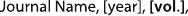}}
\fancyfoot[CE]{\vspace{-7.2pt}\hspace{-14.2cm}\includegraphics{RF}}
\fancyfoot[RO]{\footnotesize{\sffamily{1--\pageref{LastPage} ~\textbar  \hspace{2pt}\thepage}}}
\fancyfoot[LE]{\footnotesize{\sffamily{\thepage~\textbar\hspace{3.45cm} 1--\pageref{LastPage}}}}
\fancyhead{}
\renewcommand{\headrulewidth}{0pt} 
\renewcommand{\footrulewidth}{0pt}
\setlength{\arrayrulewidth}{1pt}
\setlength{\columnsep}{6.5mm}
\setlength\bibsep{1pt}

\makeatletter 
\newlength{\figrulesep} 
\setlength{\figrulesep}{0.5\textfloatsep} 

\newcommand{\topfigrule}{\vspace*{-1pt}%
\noindent{\color{cream}\rule[-\figrulesep]{\columnwidth}{1.5pt}} }

\newcommand{\botfigrule}{\vspace*{-2pt}%
\noindent{\color{cream}\rule[\figrulesep]{\columnwidth}{1.5pt}} }

\newcommand{\dblfigrule}{\vspace*{-1pt}%
\noindent{\color{cream}\rule[-\figrulesep]{\textwidth}{1.5pt}} }

\makeatother

\twocolumn[
  \begin{@twocolumnfalse}
{\includegraphics[height=0pt]{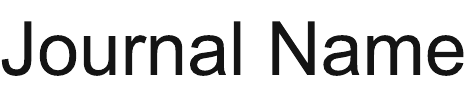}\hfill\raisebox{0pt}[0pt][0pt]
{\includegraphics[height=55pt]{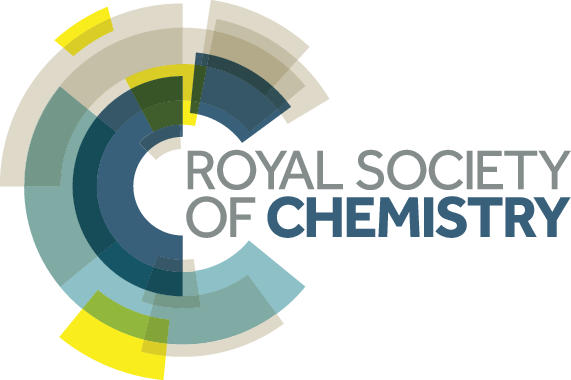}}\\[1ex]
\includegraphics[width=18.5cm]{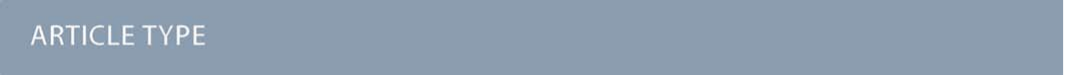}}\par
\vspace{1em}
\sffamily
\begin{tabular}{m{4.5cm} p{13.5cm} }

\includegraphics{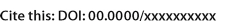} & 
\noindent\LARGE{\textbf{Depletion-induced crystallization of anisotropic triblock colloids$^\dag$}}\\

\vspace{0.3cm} & \vspace{0.3cm} \\

 & \noindent\large{Fabrizio Camerin,\textit{$^{a,b}$$^{\ast}$} Susana Marín-Aguilar,\textit{$^{a}$$^{\ast}$} and Marjolein Dijkstra\textit{$^{a,b}$}} \\

\includegraphics{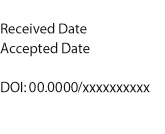} & \noindent\normalsize{The intricate interplay between colloidal particle shape and precisely engineered interaction potentials has paved the way for the discovery of unprecedented crystal structures in both two and three dimensions. Here, we make use of anisotropic triblock colloidal particles composed of two distinct  materials. The resulting surface charge heterogeneity can be exploited to generate regioselective depletion interactions and directional bonding. Using extensive molecular dynamics simulations and a dimensionality reduction analysis approach, we map out state diagrams for the self-assembly of such colloids as a function of their aspect ratio and packing fraction for varying depletant sizes in a quasi two-dimensional set-up. We observe the formation of a wide variety of crystal structures such as a herringbone, brick-wall, tilted brick-wall, and (tilted) ladder-like structures. More specifically, we determine the optimal parameters to enhance crystallization, and investigate the nucleation process.  Additionally, we explore the potential of using crystal monolayers as templates for deposition, thereby creating complex three-dimensional structures that hold promise for future applications. 
}\\

\end{tabular}

 \end{@twocolumnfalse} \vspace{0.6cm}
]

\renewcommand*\rmdefault{bch}\normalfont\upshape
\rmfamily
\section*{}
\vspace{-1cm}


\footnotetext{\textit{$^{a}$~Soft Condensed Matter \& Biophysics, Debye Institute for Nanomaterials Science, Utrecht University, Princetonplein 1, 3584 CC Utrecht, The Netherlands; $^{b}$ International Institute for Sustainability with Knotted Chiral Meta Matter (WPI-SKCM$^2$), Hiroshima University, 1-3-1 Kagamiyama, Higashi-Hiroshima, Hiroshima 739-8526, Japan;
E-mails: f.camerin@uu.nl, s.marinaguilar@uu.nl, m.dijkstra@uu.nl}}

\footnotetext{\dag~Electronic Supplementary Information (ESI) available: [details of any supplementary information available should be included here]. See DOI: 00.0000/00000000.}

\footnotetext{$\ast$~These authors contributed equally to this work}



\section{Introduction}
The availability of well-characterized colloidal suspensions has enabled us to study physical phenomena that resemble those encountered in the realm of atoms and molecules. The temporal and spatial scales inherent to colloidal systems render the investigation of these systems on a single-particle level significantly more attainable in comparison to their atomic and molecular counterparts. This accessibility has shed light on problems such as crystallization~\cite{herlach2010colloids, gasser2009crystallization} and the  nature of the glass transition~\cite{gokhale2016deconstructing, weeks2017introduction}. 
 
While originally these studies primarily involved simple hard-sphere-like colloids~\cite{pusey1987observation}, experimental techniques have increasingly advanced to have complete control over their particle shape and interactions, thereby achieving a more accurate representation of their molecular analogs. Currently, it is feasible to fabricate colloids that exhibit intricate anisotropic geometries~\cite{sacanna2011shape,hueckel2021total} and interactions~\cite{zhang2015toward, chen2011triblock,liu2020customized}. These colloids can be endowed with different surface functionalizations and directional bonding, closely emulating the principles of molecular recognition, bonding, and shape.
A notable recent example involves the self-assembly of colloidal cubic diamond structures using discrete building blocks composed of tetrahedral patchy particles coated with DNA~\cite{he2020colloidal}. Additionally, the combined use of cubic and spherical colloidal components~\cite{chakraborty2017colloidal} has facilitated the creation of colloidal molecules with precise control over molecular valency and bond flexibility, depending on the  size ratio between cubes and spheres~\cite{shelke2023flexible}.

The study of colloids has also shed light on other physical phenomena for which a similarity to the atomistic and molecular world has been found. 
An illustrative example is the investigation of the vapor-liquid interface, which can be created by adding sufficiently  large polymer to a colloidal suspension, thereby inducing a vapor-liquid phase coexistence, which is separated  by an interface~\cite{aarts2004direct}.
The driving force behind this phase separation is attributed to the depletion mechanism. This stems from the exclusion of polymers from the interstitial region  between the surfaces of two (or more) colloids which generates an osmotic pressure that brings the colloids closer together, ultimately leading to contact. Depletion thus effectively manifests as an attractive interaction between the colloids~\cite{lekkerkerker2011depletion} with its range and strength depending on the size and concentration of the added polymer chains. This type of interaction has proven to be extremely useful as a means of controlling crystallization~\cite{rossi2011cubic}.
In order to induce site-selective directional interactions for the self-assembly of intricate structures, depletion interactions have often been coupled with  lock-and-key mechanisms relying on the complementary shape of the colloids~\cite{sacanna2010lock,sacanna2011shape,wang2014three,mihut2017assembling}, or with particles exhibiting regions of different surface roughness.~\cite{kraft2012surface,wolters2015self} 

Therefore, the addition of polymers (or other depletants) to the solution serves as a valuable tool to guide the self-assembly of colloidal particles through  depletion. However, colloids  are inherently endowed with certain characteristics that contribute to their overall interaction potential.  Notably, the presence of surface charges and factors such as the  excluded volume of the colloids or their geometric shape can all be harnessed to define new assembly strategies~\cite{vogel2015advances}.
These concepts have been exploited by Liu and coworkers~\cite{liu2020tunable} who fabricated triblock colloids composed of two distinct polymeric materials and observed the formation of a remarkable array of 1D, 2D, and quasi-2D structures through scanning electron microscopy. 
These particles exhibit an ellipsoidal shape which can be described by two orthogonal axes. More specifically, these biphasic triblock colloids, fabricated by means of a so-called cluster encapsulation method~\cite{zheng2017shape,liu2020tunable}, consists of 3-(trimethoxysilyl)propyl methacrylate (TPM) and polystyrene (PS), with the former located in the center of the colloid and the latter at the tips. 
Moreover, by varying the size of the central TPM  colloid, triblock colloids with different aspect ratios can be synthesized, consequently yielding  different structures such as ladder-like chains, brick-wall configurations or herringbone patterns.

However, while this study has pioneered a novel self-assembly technique, the optimal parameters for the emergence of crystalline patterns remain elusive. For instance, it remains undisclosed whether crystallization might be more pronounced at specific colloid aspect ratios, or for different  depletant sizes, depletant concentrations, or colloid packing fractions. Furthermore, the ability of this system to form three-dimensional structures remains unexplored, a facet that could enhance its appeal from an application point of view~\cite{cai2021colloidal}.

In this work, we present a comprehensive study on the self-assembly behavior of anisotropic biphasic triblock colloids in which we aim at elucidating their behavior in forming new crystal patterns in a quasi two-dimensional set-up. To this end, we first develop a coarse-grained model that allows us to emulate the shape of the colloids as synthesized in the experiments of Ref.~\citenum{liu2020tunable}, introducing the possibility of having different interaction potentials for the different regions of the colloids and thus being able to reproduce the presence of two different materials. By tuning the size of the central particle, we can alter the aspect ratio and the patch sizes of the colloids. Using molecular dynamics simulations, we investigate the effect of packing fraction and size of the depletants on the self-assembled structures of the anisotropic triblock patchy colloids. 
We analyze such structures using a dimensionality reduction technique called principal component analysis (PCA), in which a linear combination of multiple order parameters allows the efficient identification of crystal structures in certain regions of the state diagram. 
Subsequently, the crystallization process is investigated by identifying the initial seed and following the nucleation process.
Finally, we present an outlook of our study in which we demonstrate the possibility of using triblock colloids as building blocks for the formation of novel three-dimensional crystal structures.

\section{Models and methods}

\subsection{Model of anisotropic biphasic triblock colloids}

\begin{figure}[h!]
\centering
     \includegraphics [width=0.48\textwidth]{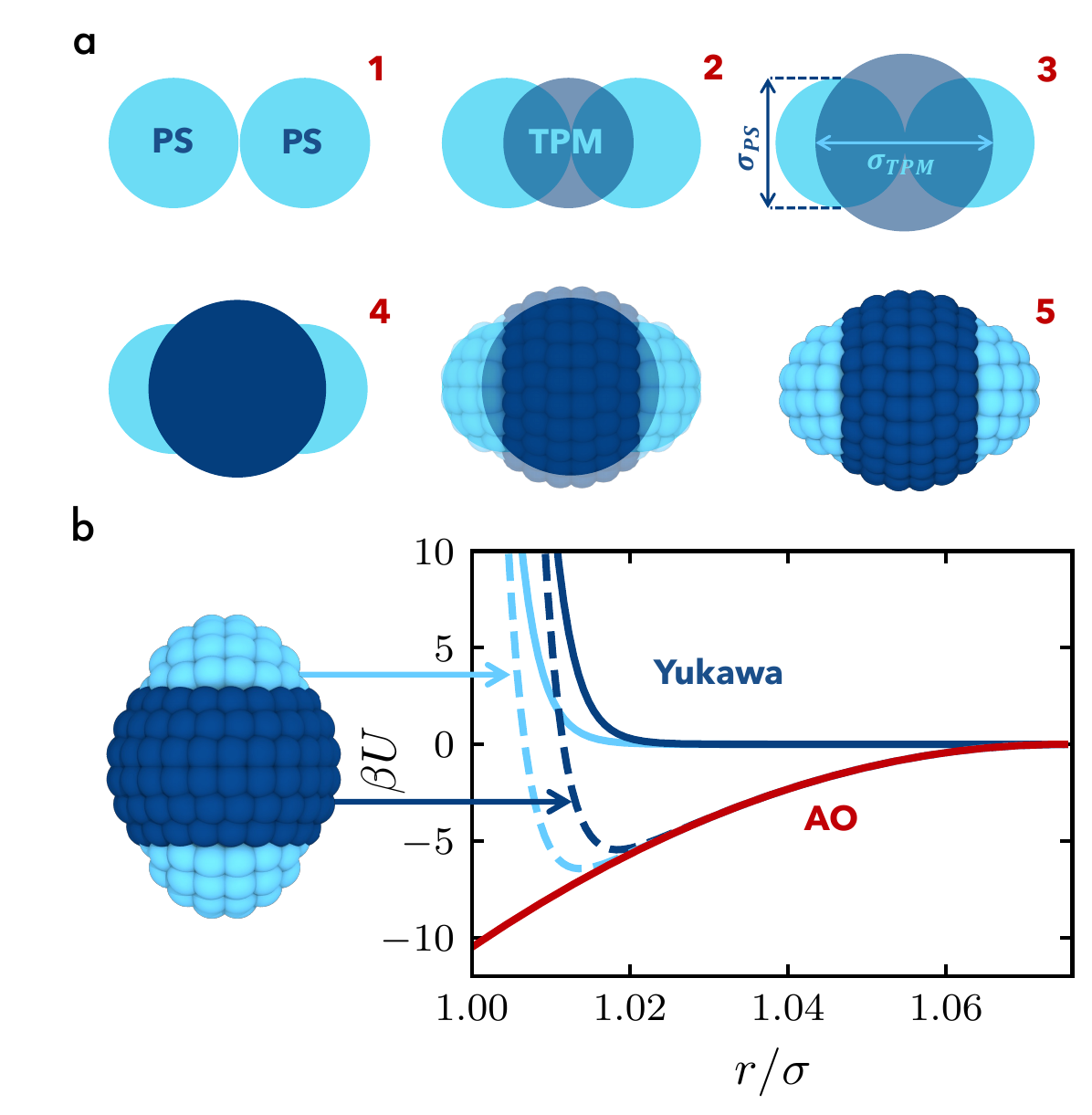} 
	 \caption{(a) Schematics showing the steps for modeling anisotropic biphasic triblock colloids involving (1) two tip spheres (PS) at contact, (2) the addition of a third central sphere (TPM), (3) the tuning of the size of the central sphere, (4) merging the surfaces of the three spherical colloids and (5) the tessellation of the surface of the resulting triblock colloid with beads. (b) Example of the overall interaction potential (dashed lines) and separate Yukawa and Asakura-Osawa (AO) interaction potential contributions (full lines) assigned to the beads depending on their position on the surface of the triblock colloid relative to the three underlying spheres (PS and TPM). Light blue is for beads belonging to the tips of the colloid, while dark blue is for those belonging to the central particle. The same Asakura-Osawa potential is assigned to all beads.
  }
	\label{fig:modelandpotential}
\end{figure} 

We develop a coarse-grained model of the anisotropic biphasic triblock colloids to mimic the experimental system presented in Ref.~\citenum{liu2020tunable}. The modelling involves the following steps, as also shown schematically in Fig.~\ref{fig:modelandpotential}(a). 
We first place two spheres of diameter $\sigma_{\rm PS}$ at contact, mimicking the PS particles in the experiments, and one sphere with diameter $\sigma_{\rm TPM}$ centered at their contact point, resembling the TPM colloid, see Fig.~\ref{fig:modelandpotential}(a1-a2). The 
shape of the biphasic triblock colloid is determined by the outer surface of the three intersecting spheres.
The aspect ratio $\lambda=2\sigma_{\rm PS}/\sigma_{\rm TPM}$ of the triblock colloids can be tuned by adjusting $\sigma_{\rm TPM}$, while we keep  the size of the outer spheres fixed at $\sigma_{\rm PS}=3.3\sigma$, where $\sigma$ is used as the unit of length in our simulations,  see Fig.~\ref{fig:modelandpotential}(a3-a4). 
We evenly tessellate the outer surface with beads of size $\sigma$ and unitary mass $m$, as illustrated in Fig.~\ref{fig:modelandpotential}(a5), following the distribution described in Ref.\citenum{deserno2004generate}. Comparable tessellation methodologies have been reported in other studies concerning patchy and Janus particles~\cite{zhang2004self, hong2006clusters}.

For each sphere of size $\sigma_{\rm PS}$, we fix the number of beads  $N_{\rm PS}=41$. The number of beads assigned to the central particle $N_{\rm TPM}$ is determined to maintain a uniform bead surface density of 
$\rho_{b}\sigma^{2}=1.2$ across all three spheres constituting the complete colloid structure.
Consequently, depending on $\lambda$, $N_{\rm TPM}$ ranges from $189$ to $51$ for $\lambda=1.30$ and $2.50$, respectively.
As the surface of the triblock  colloids corresponds to the intersection of the three spheres mentioned earlier, any overlapping beads are eliminated.
Hence, the total number of beads $N$ of the triblock particle is less than the sum of the beads initially assigned to each sphere and varies from $N=210$ for $\lambda=1.30$ to $N=158$ for $\lambda=2.50$. We verified that doubling the number of beads does not change the results other than shifting the energy scale at which the crystalline phases are observed.

\subsection{Interaction potential}
As observed in the experiments of Ref.~\citenum{liu2020tunable}, triblock colloids endowed with surface charges contingent upon the specific material, and the addition of polymer to the suspension, give rise to a complex interaction potential $\beta U^{ij}(r)$ for beads $i,j$ = TPM, PS. This can be described by the sum of a repulsive Yukawa potential~\cite{kraft2012surface,hynninen2003phase}
\begin{equation}\label{eq:y}
\beta U^{ij}_Y(r) = A^{ij} \frac{e^{-\kappa(r-\sigma)}}{r},
\end{equation}
with $A^{ij}=\beta^2 e^2 \psi_i\psi_j (\sigma_{i}\sigma_j/4\lambda_B)e^{\kappa \lbrack(\sigma_{i}+\sigma_j)/2-\sigma\rbrack}$ a prefactor that depends on the material and the size of the colloids with $\psi_{\rm PS(TPM)}$ and $\sigma_{\rm PS(TPM)}$ the zeta potential and the diameter of either the tip or central  colloid, respectively, $\lambda_B$ the Bjerrum length, $\kappa$ the inverse Debye screening length, $\beta=1/k_B T$ with $k_B$ the Boltzmann constant, $T$ the temperature, and $e$ the elementary charge, and a depletion contribution described by the Asakura-Oosawa  potential~\cite{asakura1958interaction, marin2022guiding}
\begin{eqnarray}\label{eq:ao}
\beta U_{AO}(r) &=& -\eta \left(\frac{1+q^*}{q^*}\right)^3 \times \nonumber \\
& & \left[1-\frac{3r}{2\sigma(1+q^*)}+\frac{1}{2}\left(\frac{r}{\sigma(1+q^*)}\right)^3\right],
\end{eqnarray}
where $\eta$ denotes the  packing fraction of the reservoir of depletants with which the system is in equilibrium and $q^*=\sigma_d/\sigma$ with $\sigma_d$ the size of the depletants. The depth and range of $\beta U_{AO}$ is determined by $\eta$ and $q^*$, respectively. Excluded volume interactions are accounted for by  $\beta U^{ij}_Y$. The total interaction potential
\begin{equation}
    \beta U^{ij}(r)=\begin{cases}
    \beta U^{ij}_Y(r)+\beta U_{AO}(r) & r\leq \sigma+\sigma_d \\
    0 & r > \sigma + \sigma_d,
    \end{cases}
\end{equation}
is attributed to each bead $i,j=$ TPM, PS forming the tessellated colloids. 
This approach allows us to attribute distinct interaction potentials to each bead, facilitating the modeling of various constituent materials of the colloids, such as PS or TPM as utilized in the experiments. 
The separate contributions and the total interaction potential are depicted in Fig.~\ref{fig:modelandpotential}(b), with dark blue lines for the TPM central particle and light blue lines for the PS tips.
In this work, we systematically vary the size ratio between the depletants and the PS colloids $q=\sigma_d/\sigma_{\rm PS}$  within the range of  $0.04$ to $1.2$, and the aspect ratio $\lambda$ within the range of 1.30 to 2.5, where we keep the diameter $\sigma_{\rm PS}$ of the outer sphere fixed. 
The values of $\psi_{\rm PS(TPM)},\lambda_B$ and $\kappa$ are directly derived from experimental data~\cite{liu2020tunable} and are fixed at $\psi_{\rm PS}=-0.58 k_BT/e$, $\lambda_B=2.5\times10^{-3}\sigma$, and $\kappa\sigma=114.57$, respectively (see Supplementary Information). Conversely, for beads belonging to the core particle, $\psi_{\rm TPM}=-1.36 k_BT/e$, resulting in a steeper Yukawa repulsion. Due to the small screening length $\kappa$, interactions between the substrate and the colloids  are considered negligible and therefore not accounted for.

\subsection{Simulation details}
Molecular dynamics simulations are performed using the LAMMPS package~\cite{LAMMPS}. Each triblock colloid is treated as a rigid body. For every state point, we start the simulation from two to four different initial configurations, equilibrated at a low colloid number density $\rho=N_{c}\sigma^{2}/A=0.013 $, where $N_{c}$ denotes the number of colloids and $A$ represents the area of the base of the simulation box. 
To account for the quasi two-dimensional nature of the experiments,  a gravity-like force $F_g\sigma/k_BT=0.08$ is applied to each bead in the $z$ direction of the simulation box. Unlike the $x$ and $y$ directions, periodic boundary conditions are not enforced in the $z$ direction. The magnitude of this force is such that at low and intermediate colloid densities, the longest axis of the colloids remains aligned to the $xy$ plane. 
We perform simulations in the $NVT$ ensemble for varying polymer packing fractions $0.50<\eta<1.35$ and  colloid number densities $0.020<\rho=N_{c}\sigma^{2}/A<0.030$, for  at least $t = 5\times 10^7 \delta t$ with $\delta t = 0.005 \tau$ the simulation timestep and $\tau=\sqrt{m \sigma^2/\epsilon}$ the unit of time, with $\epsilon$ setting the energy scale. We fix the reduced temperature of the system to $T^\ast=k_BT/\epsilon=3$.

To generate the training dataset for the principal component analysis (PCA), we take as initial configurations the output of the floppy box Monte Carlo method, as described below, and replicate the resulting unit cell in such a way that we obtain a crystalline configuration with a number of colloids $N_{c}>600$. To equilibrate the system, we run the simulation for $2 \times 10^{6} \delta t$. Since the specific choice of simulation parameters only marginally affects the training dataset, we fix the depletant packing fraction to $\eta=1.0$ and the depletant-to-colloid size ratio to $q=0.05$.

To study the formation of three-dimensional structures by deposition, we start from a monolayer of triblock colloids with an aspect ratio $\lambda=1.52$ arranged in a brick-wall conformation. 
This monolayer is prepared using the floppy box Monte Carlo method (see below) and the unit cell is  subsequently replicated in order to have a number of colloids $N_{c}=96$. After achieving equilibrium within the first monolayer, subsequent layers are assembled in the $z$-direction, perpendicular to the substrate, by randomly depositing additional triblock colloids in an elongated simulation box. 
In the $x$ and $y$ directions, parallel to the substrate, the box is triclinic and periodic boundary conditions are applied.
We employ the same parameters in these simulations as introduced previously for the self-assembly simulations conducted within the quasi two-dimensional setting.

\subsection{Floppy box Monte Carlo algorithm}
The floppy box Monte Carlo algorithm was originally introduced in Refs.~\citenum{filion2009efficient, de2012crystal} for predicting crystal structure candidates for colloidal particles. In the present work, we employ this method for two purposes: i) to generate the unit cells of the identified crystal phases, which enables the calculation of the order parameters for the perfect crystal structures as input for the PCA approach (see Results and discussion), and  ii) to create the colloidal monolayer, onto which we sequentially deposit additional triblock colloids for assembling three-dimensional structures.
In brief, the algorithm is based on $NPT$ Monte Carlo simulations with periodic boundary conditions. Each simulation step involves either a trial move to displace a particle or a trial move to change the volume of the system with  acceptance rules determined by the conventional Metropolis algorithm. To explore the potential arrangements of the colloidal particles, it is necessary to also consider fluctuations in the shape of the simulation box, wherein the three lattice vectors defining the box are allowed to vary independently. Consequently, a trial volume move comprises an attempt to change the orientation and the length of the lattice vectors of the simulation box. The simulations, which usually involve a small number of particles, are started in a dilute state followed by a step-wise  increase in pressure, resulting in compression of the system. In this way, the final configuration  corresponds to the unit cell of a dense crystalline structure. By running several independent simulations,  a pool of stable unit cell configurations can be obtained for triblock colloids with different aspect ratios $\lambda$. 
As compared to the original algorithm, we introduce appropriate modifications to account for the shape of triblock colloids, and for the surface tessellation of the particles through which the interaction potential is exerted. 

\subsection{Order parameters}
In order to analyze the crystalline phases resulting from the self-assembly of the triblock colloids, we employ PCA as described below in the Results and discussion section. This approach utilizes bond and orientational order parameters as input. To characterize the orientation between neighboring particles, we use the second Legendre polynomial defined as~\cite{de1993physics}
\begin{equation}
S^{j}=\frac{1}{N_{j}}\sum_k{\frac{3 \cos^2(\theta_{jk})-1}{2}}, 
\end{equation}
where the sum runs over the number of neighbors $N_j$  of particle $j$, $\theta_{jk}$ is the angle between the longest axis of the triblock colloid $j$ and the corresponding axis of the neighboring particle $k$. In addition, we define the neighbouring particles as all particles within a distance $r < 7.5\sigma$ from particle $j$. This cut-off distance has been chosen empirically to account for the first shell of nearest neighbors. For a fully isotropic phase, $S$ equals 0, while when particles are fully aligned, $S$ becomes 1.

In addition, we make use of the $n$-atic bond  order parameter~\cite{nelson2012bond}
\begin{equation} \label{eq:bond_order_parameter}
    \tilde{\Psi}_n^{j} = \frac{1}{N_j} \sum_k e^{i n  \alpha_{jk}},
\end{equation}
where  $\alpha_{jk}$ denotes the angle between the distance vector ${\bf r}_{jk}$ and the vector $(1,0)$, and $n$ governs the symmetry of the bond order parameter. 

\section{Results and discussion}
\begin{figure*}[h!]
\centering
	 \includegraphics [width=1\textwidth]{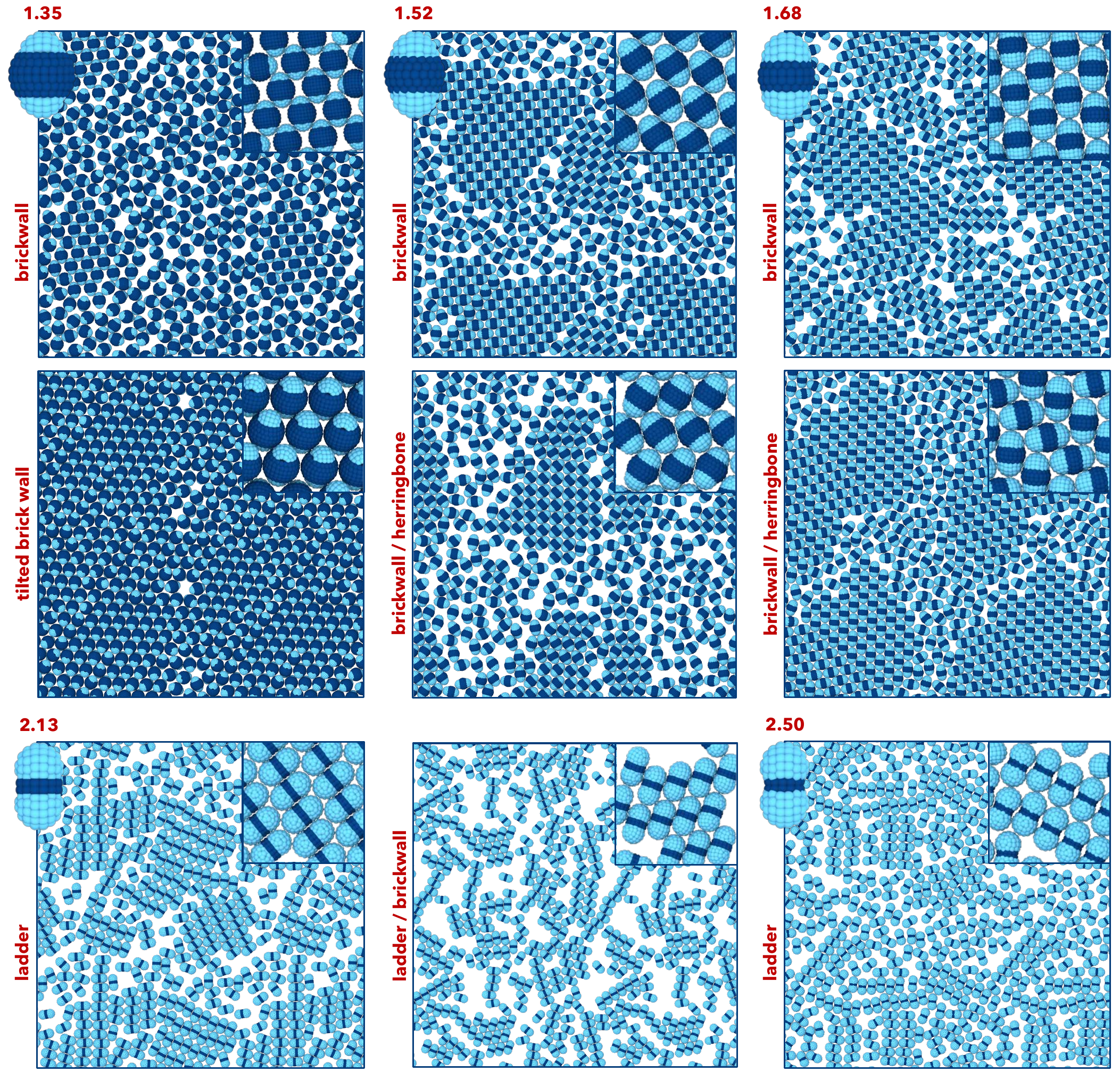} 
	 \caption{Representative simulation snapshots showing the phases that are identified for the five examined aspect ratios  $\lambda=1.35,1.52,1.68,2.13$ and $2.50$. For visual clarity, each snapshot reports a periodic image in the $x$ and $y$ directions. In each panel, we also report an enlargement of the crystal structure and a representation of the individual triblock colloid.
  }
	\label{fig:snaps}
\end{figure*} 

\subsection{Identifying crystal phases of triblock colloids}
Each triblock colloid consists of three distinct parts, with the tips composed of the same material. Similar to the experimental system~\cite{liu2020tunable}, the central part has an absolute higher surface charge, resulting in a steeper Yukawa interaction potential as compared to the tips of the colloid. The depletion interaction is the same for each bead regardless of the material type. In this way, the overall interaction potential turns out to be more attractive for the tips and slightly less attractive for the central part of the colloid. More details on the modeling and interaction potentials are provided in the Methods section and are summarized in Fig.~\ref{fig:modelandpotential}.
In addition, it is important to note that changing the aspect ratio $\lambda$ of the triblock colloid leads to changes in the surface area of the tips as well as in the central part.
For intermediate values of $\lambda$, the surface area is distributed relatively evenly between the center and the tips. However, as $\lambda$ increases further, the surface area of TPM in the center experiences a significant reduction. 
Specifically, for $\lambda > 2$, a concave colloid is created, with a smaller size of the central particle as depicted in Fig.~\ref{fig:snaps}.

Consequently, we expect the formation of different phases due to i) the different packings than can be achieved for each size ratio, and ii) the variation in exposed surface areas of the outer PS and central TPM spheres. 
To investigate this, we perform molecular dynamics simulations of triblock colloids in a quasi two-dimensional setup for varying aspect ratios  $\lambda$ along with different values for the colloid number density $\rho=N_{c}\sigma^{2}/A$, depletant packing fraction $\eta$, and depletant-to-colloid size ratio $q$, following the methodology described in the Methods section. In Fig.~\ref{fig:snaps}, we present representative simulation snapshots of the observed crystalline phases found for five different aspect ratios, $\lambda=1.35, 1.52, 1.68, 2.13$, and $2.50$. Each panel also includes an enlarged view of the corresponding phase, along with representations of the individual triblock colloids giving rise to the respective phase. 
For the smallest size ratio examined, we primarily observe the formation of two phases, namely the brick-wall and the tilted brick-wall structure. In the first case, colloids interact through their tips with all particles oriented in the same direction, and their specific shape facilitates the formation of a non-close-packed structure, often resulting in the presence of voids. 
The formation of the tilted brick-wall structure, on the other hand, arises from the lack of confinement in the third dimension. This enables  the colloids to explore the $z$-direction while remaining tied to the lower substrate. As a result, the particles lie adjacent to each other with their longest axes tilted with respect to the substrate. 
Brick-wall phases are also extensively found for $\lambda=1.52$ and $1.68$. The main difference lies in the spacing between colloids due to their different shapes, resulting in denser phases for  higher aspect ratios.
In these cases, we also observe the presence of a second phase, where each line of densely packed colloids is tilted in the opposite direction compared to the preceding one. 
Due to this characteristic, this phase is referred to as herringbone phase and is always found to be surrounded by the more dominant brick-wall phase.
In the case of the highest aspect ratios studied, namely for $\lambda=2.13$ and $2.50$, there is a preference for the triblock colloids to form linear structures, which can be seen as ladder-like configurations. These ladders can eventually join and give rise to more intricate  arrangements, resembling the brick-wall structures but oriented in a tilted manner. In the case of $\lambda=2.13$, a small nucleus of the (standard) brick-wall is also found.
From the snapshots, it is evident that the  crystalline patterns are typically not arranged in a single continuous grain but rather as separate ones with varying orientations. For the aspect ratios at which the brick-wall structure crystallizes, we also observe distinct grains that can be arranged and connected to the main one by having an orientation difference of $\approx 90^\circ$. In certain cases, we also observe that the herringbone phase is found at the boundary between two brick-wall grains with differing orientations.

After evaluating the phases that can be attained for triblock colloidal particles with varying aspect ratios, we proceed with investigating under which conditions crystallization is enhanced. The phase space is particularly large, considering that parameters like aspect ratio, colloid number density, depletant packing fraction and depletant-to-colloid size ratio all affect the self-assembled structures. Furthermore, the unique features of each phase, which could be more effectively characterized using different order parameters, present a challenge in identifying a single descriptor that universally applies to all cases.

Therefore, to facilitate the identification of phase space regions that lead to the self-assembly of crystalline patterns, we employ a dimensionality reduction technique known as  principal component analysis (PCA)~\cite{bishop2006pattern,murphy2012machine}. The idea behind PCA is to project the input data onto a reduced subset through a linear basis transformation, while retaining the most information about the data. This approach allows us to use general order parameters that are not specific to any particular phase as input, and to derive a new set of variables known as the principal components that are constructed as specific linear combinations of the original parameters. This process is designed  such  that the first few principal components account for  most of the relevant information, thus allowing the exclusion of the (higher) components with lower informational content, thereby reducing the dimensionality of the problem. A particular advantage of the PCA technique is that these principal components can be readily employed as novel order parameters with interpretable significance~\cite{van2020classifying}. 

\begin{figure}[t!]
\centering
	 \includegraphics [width=0.48\textwidth]{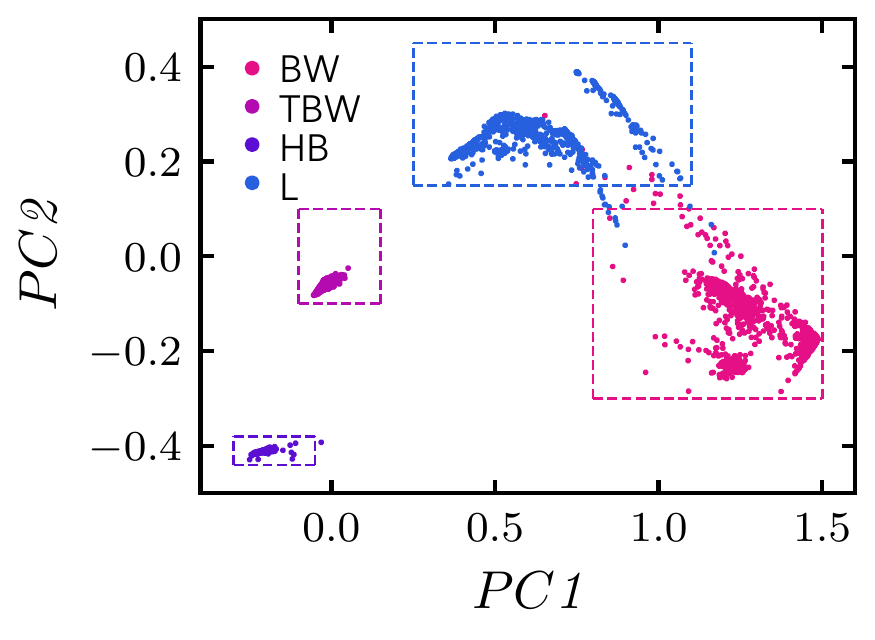}
	 \caption{Projection of the dataset with the bond  and orientational order parameters of the fully crystalline phases generated with the floppy box Monte Carlo method on the first (PC1) and second (PC2) principal components. Brick-wall (BW), tilted brick-wall (TBW), herringbone (HB) and ladder-like (L) structures are well-separated in the principal component space. First and second principal component values highlighted by the rectangular areas are  used to identify crystalline phases in self-assembly simulations.}
	\label{fig:pca}
\end{figure} 

We construct the input dataset for PCA using equilibrated, fully crystalline systems. The rationale behind this procedure is to have a representation of all crystalline phases within the principal component space. This representation is subsequently harnessed to classify  crystalline patterns in standard self-assembly simulations.
To this aim, we employ the floppy box Monte Carlo technique (see Methods) to  generate a pool of possible quasi two-dimensional crystals that can be formed by triblock colloids for varying size ratios. This pool includes fully brick-wall, tilted brick-wall, herringbone and ladder-like structures. Subsequently, these configurations are equilibrated using quasi two-dimensional molecular dynamics simulations (see Methods) which introduces thermal fluctuations to the initially perfect crystal structures.

We then characterize each of these phases by computing the second Legendre polynomials and a set of $n$-atic bond order parameters. In particular, we observe that  $n=4,6,8$, in addition to the second Legendre polynomials, offer the best representation of all the phases in this system. These order parameters thus form the input (training) dataset for the PCA analysis. 
Details regarding their definition and the way they are selected are provided in the Methods section and in the Supplementary Material, respectively. 
To ensure that each crystalline phase is adequately represented, a minimum of $N_{c}=600$ colloids per phase is employed, as described in the Methods section.
Figure~\ref{fig:pca} presents the projection of the dataset with the bond and orientational order parameters of the fully crystalline phases generated with the floppy box Monte Carlo method on the first and second  principal components ($PC1$ and $PC2$). It is immediately evident that the data points originated  from the same phase are tightly grouped within specific  regions of the principal component space. This is particularly pronounced  for the tilted brick-wall and  herringbone structures. In this way, we can define $PC1$ and $PC2$ threshold values that characterize each phase, as depicted by the dashed rectangular lines in Fig.~\ref{fig:pca}. Note that, a subtle mixing between brick-wall and ladder-like structures  occurs  around $PC1 \approx 1.0$ and $PC2 \approx 0.1$ as both structures share some orientational similarities. In any case, this small region of overlap does not hinder the identification of either phase. 

\begin{figure*}[t!]
\centering
	 \includegraphics [width=1\textwidth]{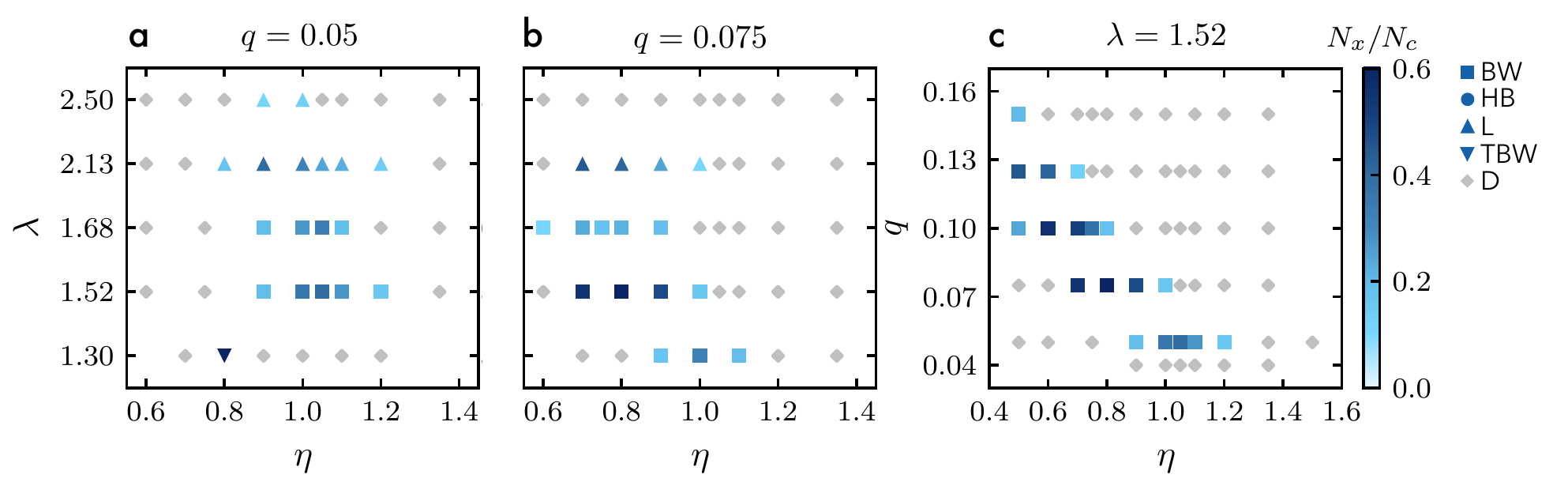} 
	 \caption{State diagram of triblock particles in  the aspect ratio $\lambda$ - depletant packing fraction $\eta$ representation for depletant-to-colloid size ratio (a) $q=0.05$ and (b) $q=0.075$. (c) State diagram of triblock particles in  the depletant-to-colloid size ratio $q$ - depletant packing fraction $\eta$ representation for the aspect ratio $\lambda=1.52$. Five different phases are identified, namely brick-wall (BW) (squares), herringbone (HB) (circles), ladder (L) (triangles), tilted brick-wall (TBW) (triangles down) and disordered (D) (diamonds) structures. For each state point, the most prevalent phase is reported, independent on the colloid number density $\rho=N_{c}\sigma^{2}/A$.  With the exception of the disordered phase, symbols are colored according to the fraction of the corresponding crystal phase identified $N_x/N_c$, in accordance with the color scale displayed in the Figure.}
	\label{fig:statediagrams}
\end{figure*} 

\subsection{State diagrams for anisotropic triblock colloids}
To investigate the self-assembly of triblock colloids, we perform molecular dynamics simulations at different colloid number densities, depletant packing fractions, and depletant sizes, for the  previously introduced aspect ratios $\lambda$. The simulations are  started from a low-density configuration in which the colloids have  random positions and orientations. Subsequently, we analyze each state point by calculating the respective principal components. This approach facilitates the identification of  different phases by comparing their principal component values to those of the fully crystalline phases (see Figure~\ref{fig:pca}). 
Depending on the number of colloids that present certain principal component values, we can quantify the prevalence of a crystal phase as $N_x/N_{c}$, where $N_x$ is the number of particles identified in a particular phase. 
We note that, unlike the fully crystalline systems,  particles in the self-assembly simulations tend to organize into distinct crystalline  grains. Consequently, particles situated at the boundaries of crystalline regions exhibit principal component values that differ from those in a fully crystallized system due to  different arrangements of their nearest neighbors. We observe that this effect may lead to underestimation of $N_x/N_{c}$, particularly  for the brick-wall configuration, while boundary regions appear to be well-captured for the other phases. To address this issue, we adopt a strategy wherein we identify a minimum of $N_{boundary}=20$ particles belonging to a brick-wall boundary. Subsequently, we compute their corresponding principal components and observe that they cluster within a specific region of  the principal component space, namely, within the range of  $0.3\leq PC1 \leq 0.38$ and $-0.1 \leq PC2 \leq 0.1$. As a result, by identifying the particles belonging to the brick-wall boundary regions, we can also classify them as part of the brick-wall phase. 

We summarize our results in the state diagrams as shown in Fig.~\ref{fig:statediagrams}(a,b), where we report for each aspect ratio $\lambda$ and depletant packing fraction $\eta$ the fraction of triblock colloids found in the most prevalent crystalline phase, as indicated by the color bar, for two depletant-to-colloid size ratios $q=0.05$ and $q=0.075$, respectively. In case the fraction of colloids belonging to the most prevalent crystalline phase is lower than $10\%$, the system is considered as disordered.
In the majority of cases, we observe minimal influence stemming from  the colloid number density $\rho=N_{c}\sigma^{2}/A$. Therefore, we present for each state point  the most prevailing phase for the entire range of colloid number densities investigated. State diagrams displaying the full dependence on $\rho$ are available in the Supplementary Information.

We  observe that the brick-wall and ladder-like structures are the two prevalent phases for values of $\lambda$ smaller and higher than 2, respectively. This suggests  that convex shapes of the triblock particles are more prone to stabilize brick-wall crystals, while the relatively smaller size of the center of the particles as compared to the tips favors the  ladder-like conformation, consistent with the patterns observed in the simulation snapshots reported in Fig.~\ref{fig:snaps}. Notably, the formation of tilted brick-wall structure is limited to a single state point for $\lambda=1.30$.
Regarding the depletant packing fraction, crystalline phases are found for $0.7 \lesssim \eta \lesssim 1.1$ for both $q$, with the most prevalent occurrences of the brick-wall structure ($N_x/N_{c}> 0.5$) at $\eta = 0.8$ for $\lambda=1.52$. In contrast, for the highest aspect ratio examined, limited crystal formation is observed, particularly for $q=0.075$ where only fully disordered states are found. We note that herringbone phases consistently nucleate alongside brick-wall phases but are less frequent in number. Consequently, they are not uncovered in the state diagrams. 

In general, we note that our method for detecting  crystalline phases may slightly underestimate the quantity of crystals within a particular phase. This issue is further enlarged by the  previously mentioned challenge of accurately identifying boundary particles. Another  aspect that has to be taken into account concerns the  selection of  the classification boundaries in the principal component space as defined in Fig.~\ref{fig:pca}. We note that a shift in these boundaries could have a small effect on the quantity of crystals. The utilization of a distinct technique for phase identification, such as one relying directly on order parameters, would likely encounter comparable issues, particularly in distinguishing different phases and accurately identifying crystal boundaries. 
Furthermore, adopting such an approach would not ensure the independent identification of phases with the use of a unique descriptor as achieved through PCA.

\begin{figure*}[t!]
\centering
\includegraphics [width=1\textwidth]{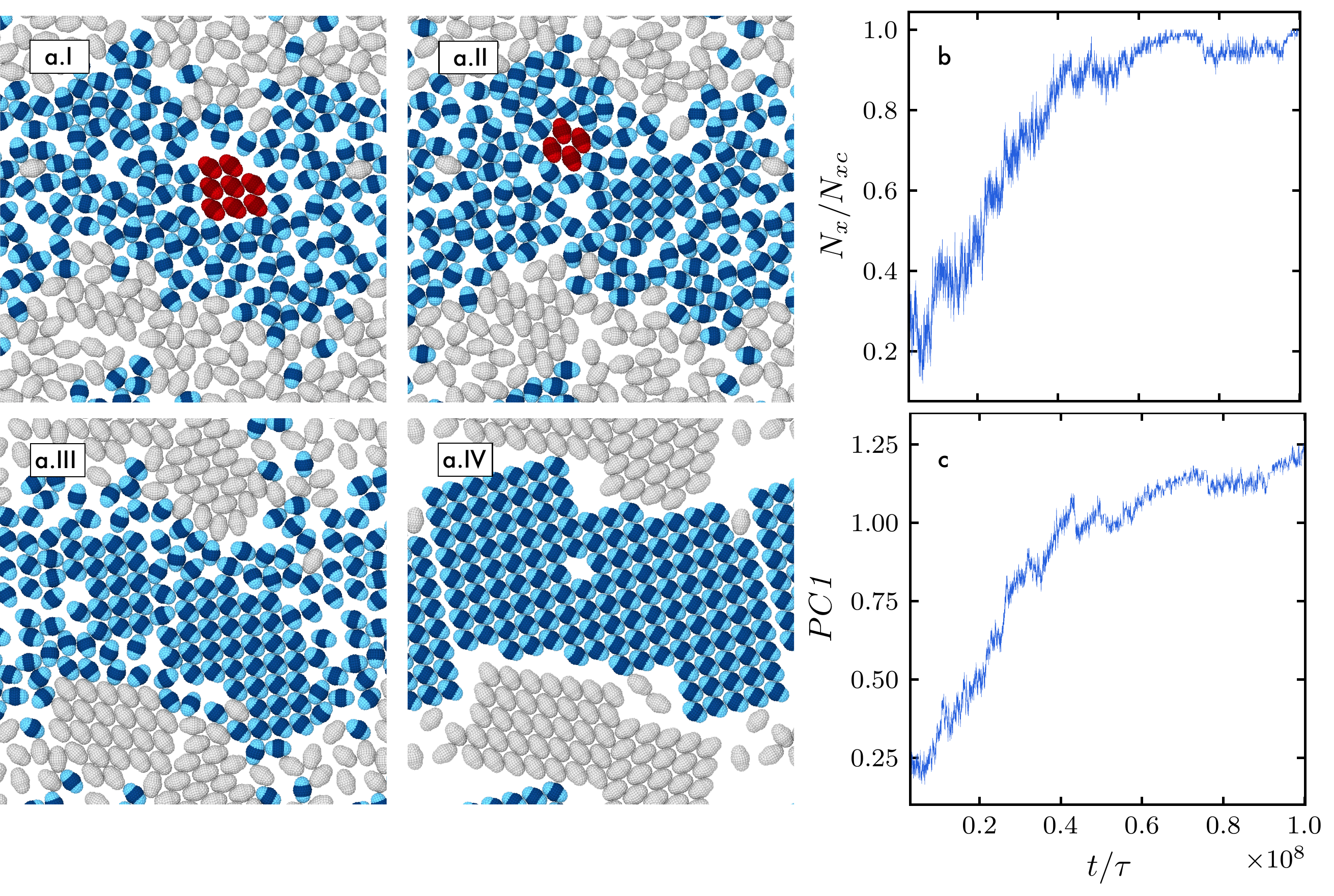} 
	 \caption{(a.I-IV) Representative time-lapse simulation snapshots showing the formation of brick-wall crystals and (b) the fraction of brick-wall crystal $N_{x}/N_{xc}$ as a function of simulation time $t/\tau$, for the largest crystal cluster  as identified in the last timestep of the simulation (blue and light blue colloids), reaching a size of  $N_{xc}$ particles. (c) The first principal component (PC1) as a function of simulation time $t/\tau$ for the same system. Data corresponds to a system of triblock colloids with an aspect ratio $\lambda=1.52$, colloid number density $\rho=N_{c}\sigma^{2}/A=0.026$, depletant-to-colloid size ratio $q=0.075$, and depletant packing fraction $\eta=0.80$. Grey colloids are not considered for the calculations  in (b) and (c). Red colloids highlight a seed from which different grains are formed in (a.I) and (a.II). 
  }
	\label{fig:nucleation}
\end{figure*}
We now focus on a specific aspect ratio, namely $\lambda=1.52$, and study in more detail the influence of the depletion interaction, in terms of  the depletant-to-colloid size ratio $q$ and  depletant packing fraction $\eta$. The corresponding state diagram is reported in Fig.~\ref{fig:statediagrams}(c). Once again, we show the largest value of the prevailing phase encountered across the range of colloid densities explored. We note that the formation of brick-wall crystals emerges as a result of the intricate interplay between $q$ and $\eta$. This relationship entails that a reduction in depletant concentration corresponds to  an increase in depletant size. Given that  these two variables control the range and  depth of the depletion potential, it becomes evident that crystal nucleation requires a specific range and strength of the attraction potential. In this way, by meticulously tuning both the range and concentration of  depletants, it becomes possible to achieve substantial degree of crystal formation. Alternatively, a different choice of the parameters would instead lead to the formation of a disordered phase, as indicated by the grey diamond symbols in the state diagram. It is pertinent to observe that different types of disordered structures are found for different values of depletant size $q$. 
 Specifically, when dealing with larger  attraction ranges, gel-like structures emerge on the substrate. These structures are typically characterized by random arrangements of colloids forming a percolating network. In some cases, we also observe the formation of networks that exhibit a certain degree of crystalline order, although this order is less pronounced than under optimal conditions for the formation of crystal grains. The self-assembly of colloids into crystalline networks have been previously reported for other systems with short-range attractions~\cite{marin2022guiding,fortini2008crystallization,zaccarelli2008gelation}. Representative snapshots depicting  such phases are shown in the Supplementary Information for systems with a fixed colloid number density and depletant packing fraction, while varying the depletant size $q$.

\subsection{Nucleation of crystalline phases}
In order to investigate the crystal formation process, we closely monitor the nucleation at specific state points. To this end, we present a sequence of simulation snapshots in Fig.~\ref{fig:nucleation}(a), showcasing the progressive  formation of a brick-wall crystal of  triblock colloids with an aspect ratio of  $\lambda=1.52$, depletant-to-colloid size ratio $q=0.075$, depletant packing fraction $\eta=0.80$, and colloid number density $\rho=0.026$.  
In all frames, the colloids forming the largest crystal cluster at the end of the simulation are highlighted in shades of blue and light blue. The identification of the latter is based on the criterion according to which only colloids with a  second Legendre polynomial value $S>0.8$ are considered. 
In this analysis, grey colloids are excluded as they only contribute to the formation of smaller crystallites.

The fraction of crystal that is formed over time within the largest cluster $N_x/N_{xc}$ is reported in Fig.~\ref{fig:nucleation}(b), where $N_{xc}$ corresponds to the total number of particles that constitute the largest crystallite at the end of the simulation. 
We observe that the formation of the brick-wall structure in the final timestep results from the merging of two distinct crystals, both originating  from different seed structures. These seeds can already be identified in the initial two simulation snapshots as highlighted in red, each occurring within the initial $10\%$ of the  simulation run. In both cases, crystallization occurs through the formation of a grain wherein a small cluster of particles align to form a nucleus, around which additional  colloids assemble. As compared to the analysis performed experimentally in Ref.~\citenum{liu2020tunable}, the seeds observed within the simulations  correspond to the formation of what is termed a `superseed'. 
In many cases, we also observe the emergence of precursor assemblies involving three or more colloids; however these configurations are short-lived and do  not persist over time (not shown). In this specific case, the two distinct grains merge at approximately $0.2\times10{^8}$ $t/\tau$. Subsequently, the grain steadily grows until reaching  a plateau.

\begin{figure*}[t!]
\centering
      \includegraphics [width=1.08\textwidth]{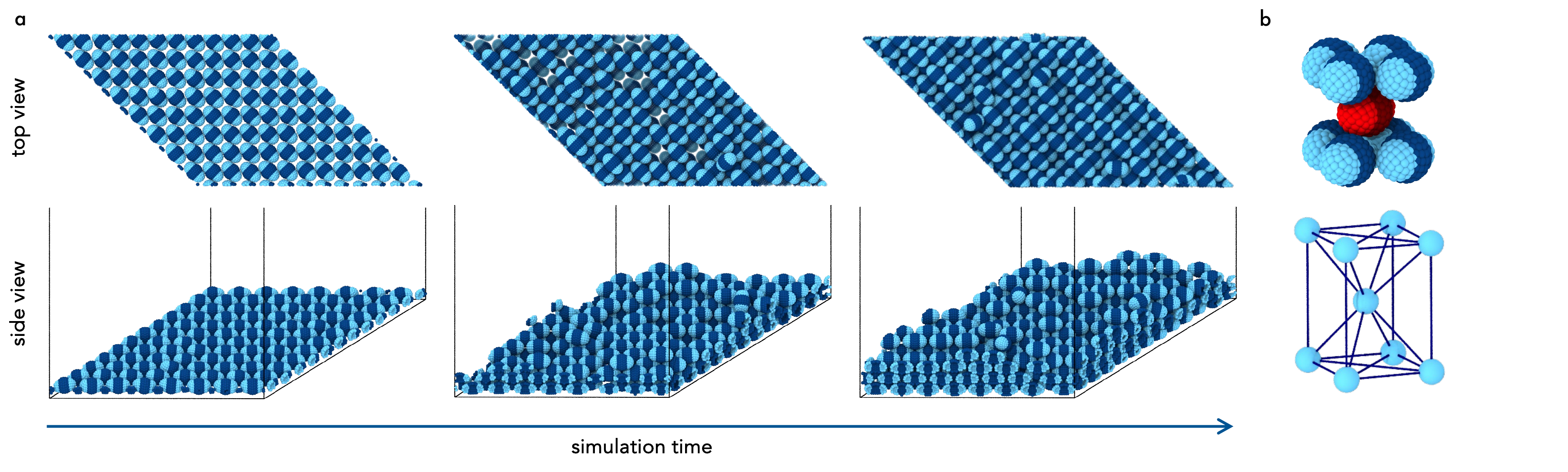} 
	 \caption{(a) Representative simulation snapshots showing the deposition of triblock colloids onto a brick-wall crystalline monolayer as a function of simulation time (from left to right). The top (bottom) panels show the top (lateral) view of the elongated simulation box.
     (b) Body-centered tetragonal (BCT) unit cell formed by the three-dimensional crystal. The upper figure shows the arrangement of the particles in their unit cell and the bottom one the corresponding centers of mass.}
	\label{fig:deposition}
\end{figure*} 

Furthermore, we can follow the progression of crystal formation by utilizing  the new order parameter derived from the principal component analysis. In Figure~\ref{fig:nucleation}(c), we report the value of the first principal component $PC1$ as a function of simulation time for the corresponding run. As expected, the trend for $PC1$ closely resembles that of $N_x/N_{xc}$, demonstrating that $PC1$ can indeed be exploited as a new order parameter for monitoring colloid crystallization. Further insight into the nucleation behavior for different simulation runs with the same parameters is provided  in the Supplementary information. 

In all cases, we observe that the largest crystal is obtained at approximately $60\%$ of the simulation duration, after which only a small number of  colloids join  the primary crystal. Similar to the first case we examined, this crystal is formed through the merging of distinct crystallites that grow and reorient over time. The formation of the herringbone structure, also reported in the Supplementary Information, is facilitated by the presence of pre-existing  brick-wall structures and is initiated by a seed of four particles, arranged pairwise with two  particles adjacent to each other.

\subsection{Formation of three-dimensional structures}
All the structures we have analyzed so far exhibit quasi two-dimensional characteristics, where  the extension in the third dimension is solely given by the volume of the triblock colloids. 
At a fundamental level, the formation of these structures show the potential to select the material and  the shape of the colloids for  directed self-assembly. However,  from an application perspective the use of such colloids for the formation of three-dimensional structures is particularly intriguing. These structures could hold promise, for instance, as colloidal photonic crystals~\cite{cai2021colloidal}. To explore this prospect, we perform molecular dynamics simulations that  exploit the deposition of colloids on a substrate. 

Specifically, our focus lies on triblock colloids with an aspect ratio of $\lambda=1.52$. We initiate the process by forming a brick-wall crystalline monolayer, and by subsequently depositing additional colloids within an elongated simulation box. Further details on these simulations are reported in the Methods section.
The outcomes of our simulations are shown in Fig.~\ref{fig:deposition}(a) in the form of representative simulation snapshots as a function of simulation time. The starting point is indeed the creation of a crystalline monolayer that serves as a template for the subsequent layers. The subsequent colloids are deposited in the empty spaces left by the former layer in such a way that an alternating stacking pattern is created. 
Overall, the structure maintains a body-centered tetragonal (BCT) arrangement as shown in Fig.~\ref{fig:deposition}(b).

The formation of crystalline order in the third dimension is confirmed by computing the principal components as a function of the $z$-direction, obtaining per layer values of $PC1\approx1.45$ and $PC2\approx-0.11$ (see Figure~\ref{fig:pca}). We underline that the deposition procedure is performed in such a way that individual colloids can progressively arrange on those previously deposited. We verified that faster deposition rates typically tend to result in the formation of disordered states without the ability to  restore ordered arrangements at longer times. 

This preliminary study on the formation of three-dimensional structures through deposition can  be extended to other size ratios for which, depending on the first template layer, different structural arrangements  can be anticipated. Additionally, it will  be of interest to assess the experimental feasibility and reproducibility of the formation of such lattices.

\section{Conclusions}
The synergistic use of shape anisotropy and complex interaction potentials facilitates the assembly of intricate  crystal phases. In particular, the use of depletion interactions can play a pivotal role, as it offers effective  control over the range and strength of the attractive forces.  Additionally, the  implementation of such interactions within an experimental framework is relatively straightforward~\cite{rossi2015shape, feng2015re}. 

Inspired by the experimental study by Liu and coworkers~\cite{liu2020tunable}, we have introduced a model for investigating  the self-assembly behavior of  biphasic triblock particles through molecular dynamics simulations. In a quasi two-dimensional setting, we have observed the emergence of different crystal phases  depending on the aspect ratio of the colloids. These structures typically initiate from different seeds and subsequently merge during the course of the simulations. However, in most cases, we observed the formation of multiple crystallites with varying  orientations, which persist over time. These observation are  consistent with the bright-field images reported experimentally~\cite{liu2020tunable}.

To classify the phases and  determine optimal conditions for crystallization, we exploited PCA  as a dimensionality reduction technique. This analysis approach proved to be very effective in discerning  crystalline phases through the combination of multiple non-specific bond order parameters. Consequently, we found that brick-wall structures are markedly prevalent for aspect ratios $\lambda<2$, while ladder-like structures dominate the state diagrams for higher aspect ratios. The formation of crystalline patterns occurs in specific areas of the parameter space, intricately controlled by the size and the packing fraction of the depletants in the systems. Beyond this range, the colloids form disordered phases, with a preference for disordered percolating structures when the depletants are of substantial size. In addition, we have demonstrated the capacity of these colloids to assemble three-dimensional structures through a deposition process. 

The range of aspect ratios with which individual building blocks can be synthesized offers the potential to generate structures with a wide array of  properties and characteristics. These colloidal crystals hold promise for various applications, including harnessing their optical properties or utilizing them as model systems to study phenomena revealed by molecular systems. 
We hope that our study stimulates further experimental exploration into  this class of colloids, along with a clever combination of different materials on particles of varying shapes. The strategic exploitation of depletant interactions holds potential for creating and investigating novel structures with desired characteristics.

\section*{Author Contributions}
Author contributions are defined based on the CRediT (Contributor Roles Taxonomy). Conceptualization: F.C., S.M.-A., M.D.; Formal analysis: F.C., S.M.-A.; Funding acquisition: M.D.; Investigation: F.C., S.M.-A.; Methodology: F.C., S.M.-A., M.D.; Project administration: M.D.; Software: F.C., S.M.-A.; Supervision: M.D.; Validation: F.C., S.M.-A.; Visualization: F.C., S.M.-A.; Writing – original draft: F.C., S.M.-A., M.D.; Writing – review and editing: F.C., S.M.-A., M.D..

\section*{Conflicts of interest}
There are no conflicts to declare.

\section*{Acknowledgements}
We thank Frank Smallenburg for sharing the floppy box Monte Carlo algorithm code and David J. Pine for fruitful discussions. The authors acknowledge funding from the European Research Council (ERC) under the European Union's Horizon 2020 research and innovation program (Grant agreement No. ERC-2019-ADG 884902, SoftML). F.C. and M.D. acknowledge funding from the World Premiere International (WPI) Research Center Initiative of the Japanese Ministry of Education, Culture, Sports, Science and Technology (MEXT).



\balance
\clearpage
\newpage
\twocolumn[
  \begin{@twocolumnfalse}
{\includegraphics[height=0pt]{journal_name}\hfill\raisebox{0pt}[0pt][0pt]
{\includegraphics[height=55pt]{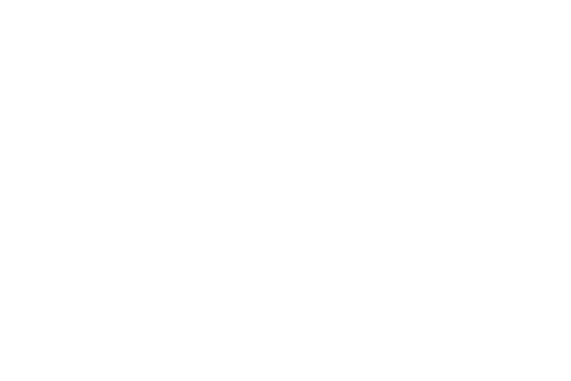}}\\[1ex]
\includegraphics[width=18.5cm]{header_bar}}\par
\vspace{1em}
\sffamily
\begin{tabular}{m{4.5cm} p{13.5cm} }

\includegraphics{DOI} & \noindent\LARGE{\textbf{Depletion-induced crystallization of anisotropic triblock colloids$^\dag$}} \\

& \medskip \noindent\LARGE{\textbf{Supplementary Information}} \\

\vspace{0.3cm} & \vspace{0.3cm} \\

 & \noindent\large{Fabrizio Camerin,\textit{$^{a,b}$$^{\ast}$} Susana Marín-Aguilar,\textit{$^{a}$$^{\ast}$} and Marjolein Dijkstra\textit{$^{a,b}$}} \\

\end{tabular}

 \end{@twocolumnfalse} \vspace{0.6cm}

]

\renewcommand*\rmdefault{bch}\normalfont\upshape
\rmfamily
\section*{}
\vspace{-1cm}


\footnotetext{\textit{$^{a}$~Soft Condensed Matter \& Biophysics, Debye Institute for Nanomaterials Science, Utrecht University, Princetonplein 1, 3584 CC Utrecht, The Netherlands; $^{b}$ International Institute for Sustainability with Knotted Chiral Meta Matter (WPI-SKCM$^2$), Hiroshima University, 1-3-1 Kagamiyama, Higashi-Hiroshima, Hiroshima 739-8526, Japan;
E-mails: f.camerin@uu.nl, s.marinaguilar@uu.nl, m.dijkstra@uu.nl}}


\footnotetext{$\ast$~These authors contributed equally to this work}


\renewcommand{\theequation}{S\arabic{equation}}\setcounter{equation}{0}
\renewcommand{\thefigure}{S\arabic{figure}}\setcounter{figure}{0}
\renewcommand{\thetable}{S\arabic{table}}\setcounter{table}{0}

\section*{Simulation parameters}
We present  the experimental values for the inverse Debye screening length $\kappa$, the Bjerrum length $\lambda_B$, and the zeta potential for PS and TPM, $\psi_{PS}$ and $\psi_{TPM}$, respectively, along with their conversions to simulation units in Table~\ref{tab:parameter}.

\begin{center}
\begin{table}[ht]
\begin{tabular}{ | c |c| c|  } 
  \hline 
  Parameter  & Experimental units & Simulation units \\  
  \hline \hline
  $\kappa$  & $0.4$ nm$^{-1}$ & $114.57 \sigma^{-1}$\\ 
  $\lambda_B$ & $0.71$nm & $2.5\times10^{-3}\sigma$\\ 
  $\psi_{PS}$ & $-15~$mV &$-0.58 k_BT/e$ \\ 
  $\psi_{TPM}$ & $-35~$mV &$-1.36 k_BT/e$ \\
  \hline
\end{tabular}
\caption{Parameters used in the Yukawa potential (see Eq. 1 in the main text) for the two types of materials, PS and TPM, in experimental units retrieved from Ref.~\citenum{liu2020tunable} and converted to simulation units: $\kappa$ the inverse Debye screening length, $\lambda_B$ the Bjerrum length, $\psi_{PS}$ the zeta potential of polystyrene (PS) located at the tips of the triblock colloid and $\psi_{TPM}$ the zeta potential of 3-(trimethoxysilyl)propyl methacrylate (TPM) located at the center of the triblock colloids.}
\label{tab:parameter}
\end{table}
\end{center}

\section*{Principal component analysis: selection of the input order parameters}

\begin{figure*}[ht!]
\centering
	 \includegraphics [width=1\textwidth]{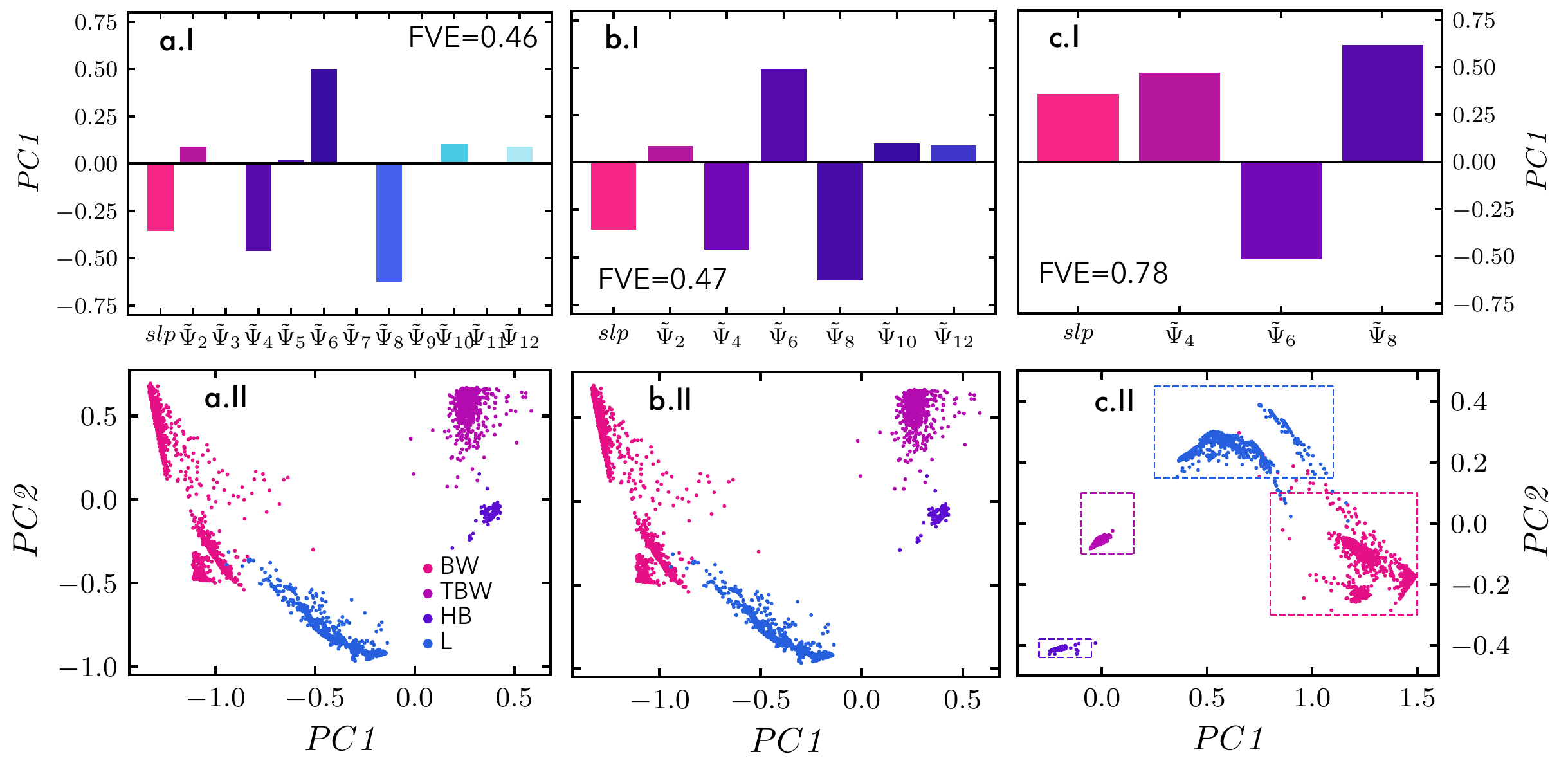} 
	 \caption{(Top, I) First principal component (PC1) coefficients and fraction of variance explained (FVE). (bottom, II) Projection of the dataset with  (a) all $n$-atic bond parameters $\Psi$ from $n\in [2,12]$ and the second Legendre polynomials (slp), (b) only the even $n$-atic bond order parameters and the second Legendre polynomials, and (c) $\tilde{\Psi}_4$, $\tilde{\Psi}_6$, $\tilde{\Psi}_8$ and the second Legendre polynomials  of fully crystalline configurations onto the first principal component (PC1) and second principal component (PC2) plane.}
	\label{fig:pca_others}
\end{figure*}

To classify the different phases found for systems of triblock colloids, we employ principal component analysis (PCA). Here, we use as input the information captured by the $n$-atic bond order parameters and the second Legendre polynomials of configurations of fully crystallized systems of triblock colloids for varying aspect ratios $\lambda$. To determine the most relevant features that describe all the phases, we first apply PCA to a large set of parameters and discard those that contribute less to the first principal component, based on their first principal component coefficients. We then apply once more the PCA to the remaining features and we repeat the procedure. Here, we start with a set of all $n$-atic bond order parameters from $n\in [2,12]$ in addition to the second Legendre polynomials. We show in Fig.~\ref{fig:pca_others}(a) the coefficients for each of the order parameters and the corresponding first and second principal component scatter plot of all the phases. We observe that the contribution of each odd $n$-atic bond order parameter tends towards zero. Hence, we discard them as relevant features. In the next step, we compute the PCA only for the even $n$-atic bond order parameters and the second Legendre polynomials as shown in Fig.~\ref{fig:pca_others}(b). We note that this step yields results very similar to the ones from the previous step, since the coefficients of all the odd $n$-atic bond order parameters were already nearly zero. Nevertheless, certain features, particularly $\tilde{\Psi}_4$, $\tilde{\Psi}_6$, $\tilde{\Psi}_8$ as well as the second Legendre polynomial,  contribute stronger to the first principal component. Subsequently, we conduct a final PCA solely using these features, and the corresponding results are shown in Fig.~\ref{fig:pca_others}(c). By applying this approach, we successfully maximize the fraction of explained variance for the first principal component to $FVE=0.78$ and for the second to $FVE=0.12$. This means that by using the first two principal components, we capture $90\%$ of the information carried by the various phases. This efficacy is reflected in the scatter plot shown in Fig.~\ref{fig:pca_others}(c.II),  where all the phases are well separated within the principal component space.

\section*{State diagrams of triblock colloids in the colloid number density - depletant packing fraction representation}
\begin{figure}[ht]
\centering
	 \includegraphics [width=0.5\textwidth]{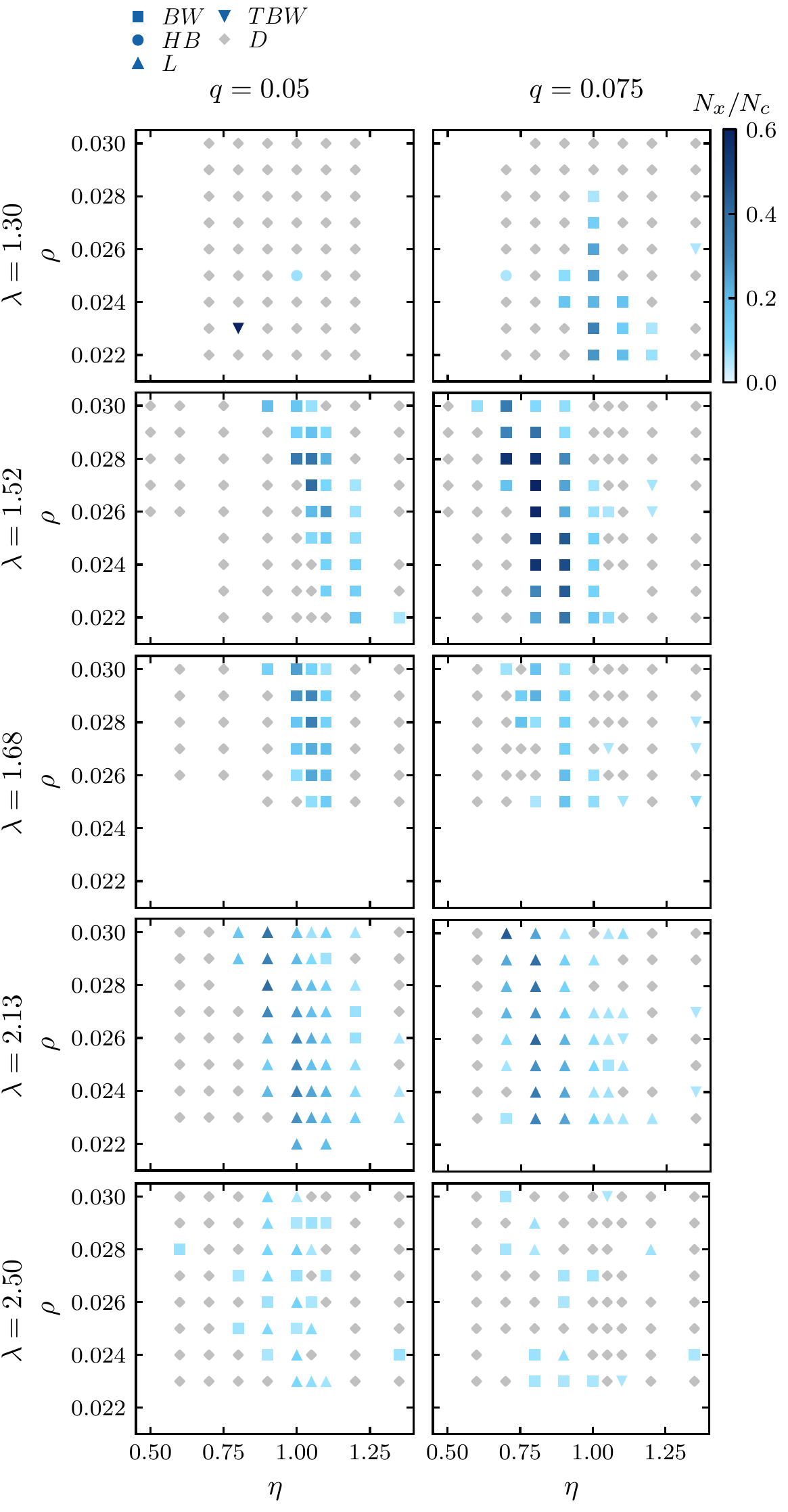} 
	 \caption{State diagrams of  triblock  colloids with varying aspect ratio $\lambda$ in  the colloid number density $\rho=N_c\sigma^2/A$ -  depletant packing fraction $\eta$ representation for depletant-to-colloid size ratio (left) $q=0.05$ and  (right) $q=0.075$. Five different phases are identified, namely brick-wall (BW) (squares), herringbone (HB) (circles), ladder-like (L) (triangles), tilted brick-wall (TBW) (triangles down) and disordered (D) (diamonds) structures. For each state point, the most prevalent phase is reported. Except for the disordered phase, symbols are assigned colors according to the maximum crystal fraction $N_x/N_{c}$, following  the color scale displayed in the Figure.
  }
	\label{fig:statediagramdensity}
\end{figure}

In Fig.~\ref{fig:statediagramdensity}(a,b), we present the state diagrams of  triblock  colloids with varying aspect ratio $\lambda$ in  the colloidal number density $\rho=N_{c}\sigma^{2}/A$ -  depletant packing fraction $\eta$ representation for two different depletant-to-colloid size ratios $q=\sigma_d/\sigma_{PS}$. This analysis allows us to determine the maximum fraction of crystalline structures for each $\eta$, as reported in Fig. 4(a,b) of the main manuscript.

\begin{figure*}[h!]
\centering
	 \includegraphics [width=0.9\textwidth]{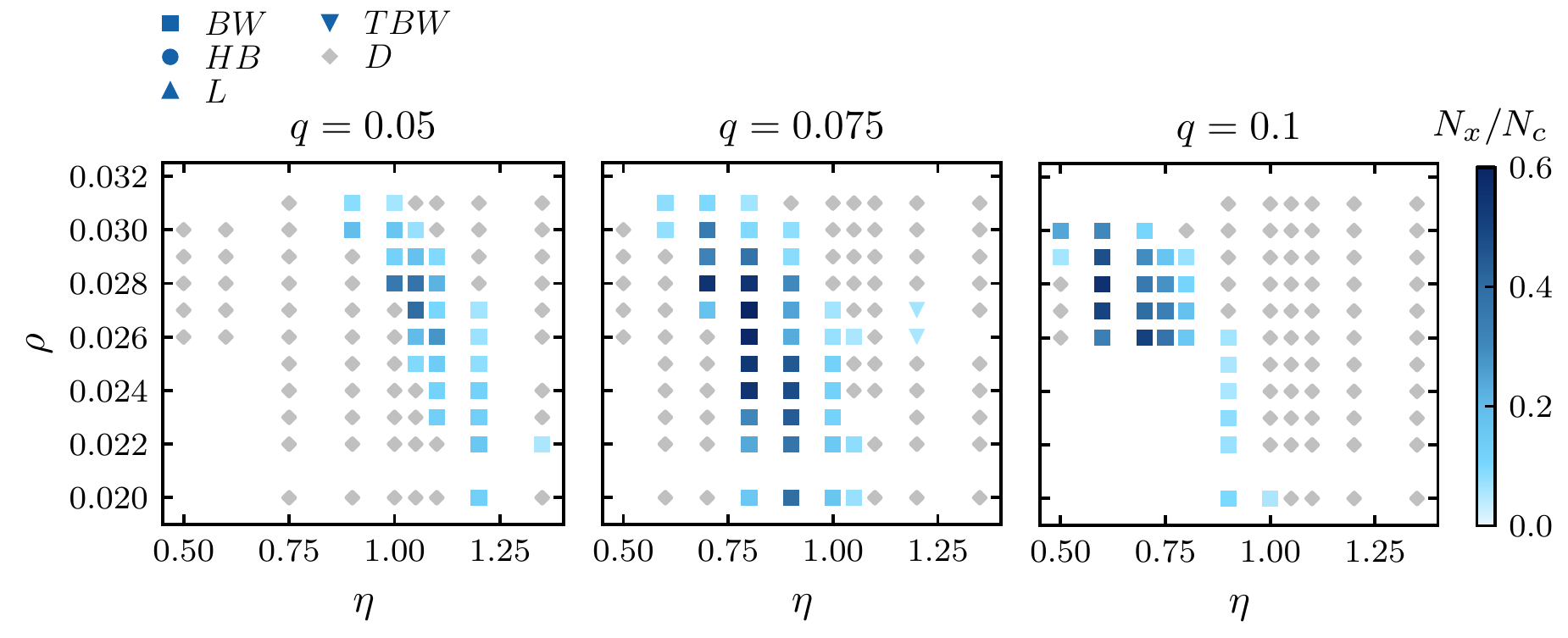} 
	 \caption{State diagrams of  triblock  colloids with varying aspect ratio $\lambda=1.52$ in  the colloid number density $\rho=N_c\sigma^2/A$ -  depletant packing fraction $\eta$ representation for depletant-to-colloid size ratio $q=0.05, 0.075,$ and $0.1$ (from left to right). Five different phases are identified, namely brick-wall (BW) (squares), herringbone (HB) (circles), ladder-like (L) (triangles), tilted brick-wall (TBW) (triangles down) and disordered (D) (diamonds) structures.  For each state point, the most prevalent phase is reported. Except for the disordered phase, symbols are assigned colors according to the maximum crystal fraction $N_x/N_{c}$, following the color scale displayed in the Figure.
  }
	 \label{fig:statediagram_1.52}
\end{figure*}

In Fig.~\ref{fig:statediagram_1.52}, we show the state diagram of triblock colloids with an  aspect ratio $\lambda=1.52$ in the colloid number density $\rho=N_{c}\sigma^{2}/A$ -  depletant packing fraction $\eta$ representation for  three different depletant-to-colloid size ratios $q=0.05$, $q=0.075$, and $q=0.1$. This data is used to extract the maximum fraction of crystal for each depletant packing fraction $\eta$ and depletant-to-colloid size ratio $q$   as shown in Fig. 4(c).

\section*{Nucleation}
\begin{figure*}[th!]
\centering
	 \includegraphics [width=0.9\textwidth]{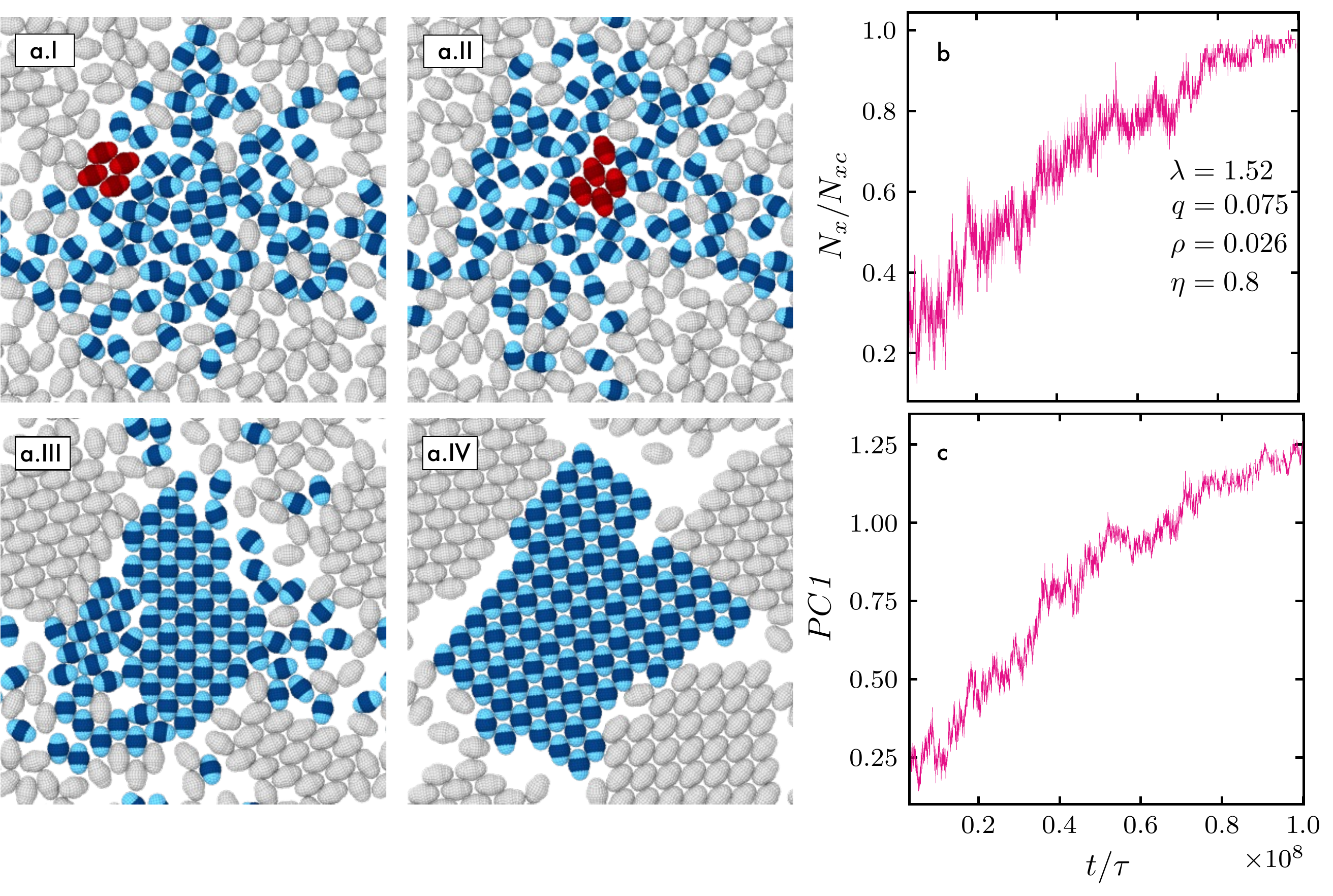} 
	 \caption{(a.I-IV) Representative time-lapse simulation snapshots showing the formation of brick-wall crystals, (b) the fraction of brick-wall crystal $N_x/N_{xc}$ as a function of simulation time $t/\tau$, and  (c) the first principal component (PC1) as a function of simulation time $t/\tau$ in a system of triblock  colloids with aspect ratio $\lambda=1.52$, depletant-to-colloid size ratio $q=0.075$, depletant packing fraction $\eta=0.8$ and colloid number density $\rho=N_{c}\sigma^{2}/A=0.026$ for the largest crystal grain identified in the last timestep of the simulation, reaching a size of $N_{xc}$ particles. Red colloids highlight seeds of two grains that will be part of the largest crystal grain at the last timestep of the simulation. Grey colloids are not considered for the calculation of $N_x/N_{xc}$ in (b) and for the PC1 in (c). 
  }
	\label{fig:nucleation_q0.075}
\end{figure*}

To investigate the nucleation behavior of the phases under examination, we conduct multiple simulation runs, allowing us to study various nucleation pathways for the same set of parameters. In particular, we focus on the aspect ratio $\lambda=1.52$ and depletant-to-colloid size ratios $q=0.05$ and $q=0.075$, which yield the highest fraction of particles in a crystalline environment. 
In Fig. 5 of the main manuscript, we present one of the simulation runs corresponding to $q=0.075$ and $\eta=0.8$. Here, we report an additional simulation with the same parameters in Fig.~\ref{fig:nucleation_q0.075}. Comparing the two simulations,  we  observe analogous nucleation behavior. Initially,  a small crystallite emerges, serving as a seed for the alignment of the surrounding particles, leading to the growth of a larger crystal grain. This progressive increase in the size of the crystal grain over time is evident from  the time evolution of the fraction of particles in the largest crystal grain and the corresponding principal component, as shown in Fig.~\ref{fig:nucleation_q0.075}(b) and (c).

\begin{figure*}[ht]
\centering
	 \includegraphics [width=0.9\textwidth]{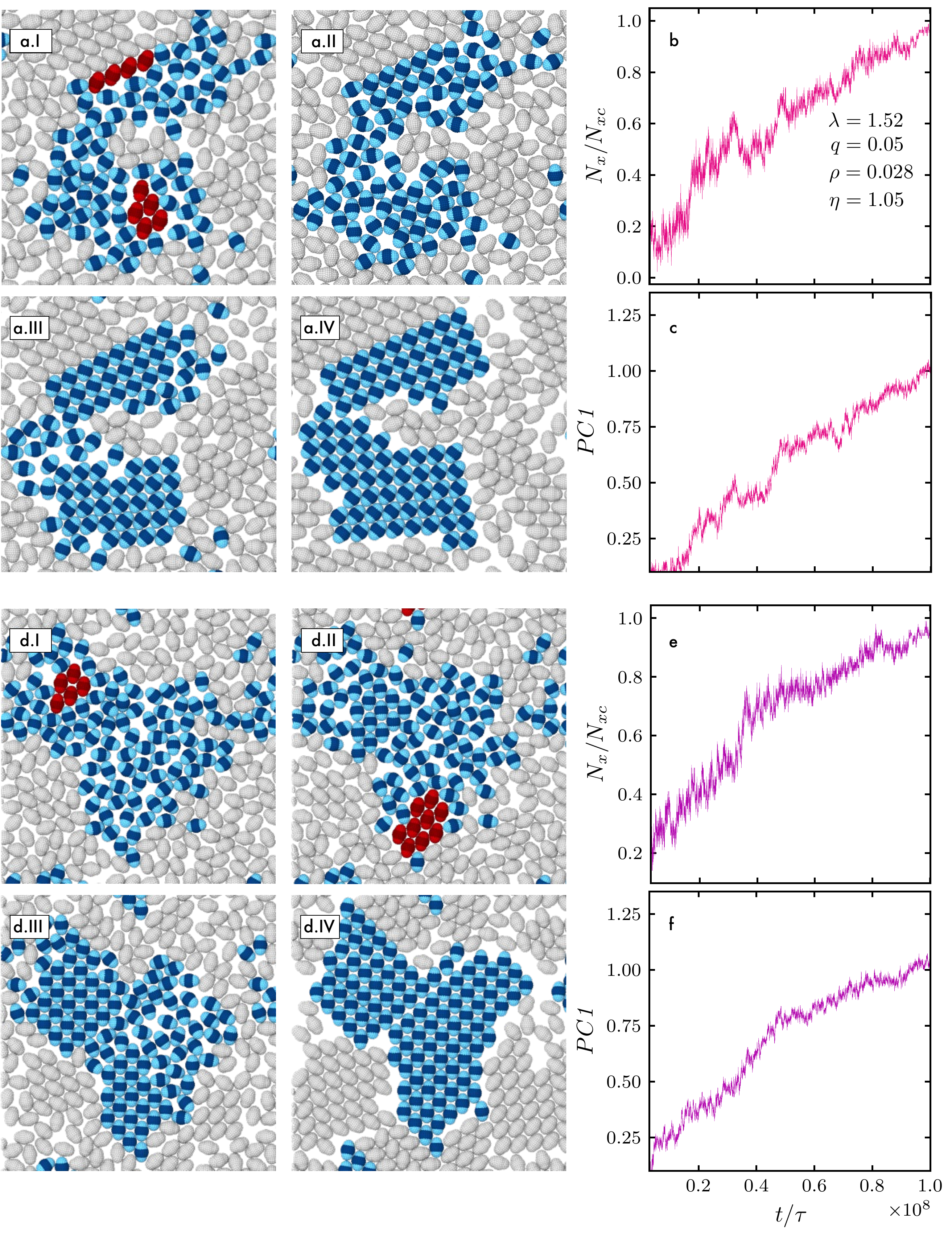} 
	 \caption{(a.I-IV, d.I-IV) Representative time-lapse simulation snapshots showing the formation of brick-wall crystals, (b,e) the fraction of brick-wall crystal $N_x/N_{xc}$ as a function of simulation time $t/\tau$, and (c,f) the first principal component (PC1) as a function of simulation time $t/\tau$ for two independent simulation runs, respectively, in a system of triblock colloids with aspect ratio $\lambda=1.52$, depletant-to-colloid size ratio $q=0.05$, depletant packing fraction $\eta=1.05$ and colloid number density of $\rho=N_c\sigma^2/A=0.028$ for the largest crystal grain identified in the last timestep of the simulation, reaching a size of $N_{xc}$ particles. Red colloids highlight the seeds that will be part of the largest crystal grain at the last timestep of the simulation. Grey colloids are not considered for the calculation of $N_x/N_{xc}$ in (b,c) and for the PC1  in (e,f).}
	\label{fig:nucleation_q0.05}
\end{figure*}

We extend our investigation to the nucleation of a system with $q=0.05$, as illustrated in Fig.~\ref{fig:nucleation_q0.05}. Once again, we observe a similar behavior, with the formation of crystal grains that, after $40\%$ of the total simulation time, merge leading to the formation of a larger crystal structure. This phenomenon is captured by a steeper increase in the fraction of particles in the largest crystal grain and the corresponding first principal component.

\begin{figure*}[hb!]
\centering
	 \includegraphics [width=0.5\linewidth]{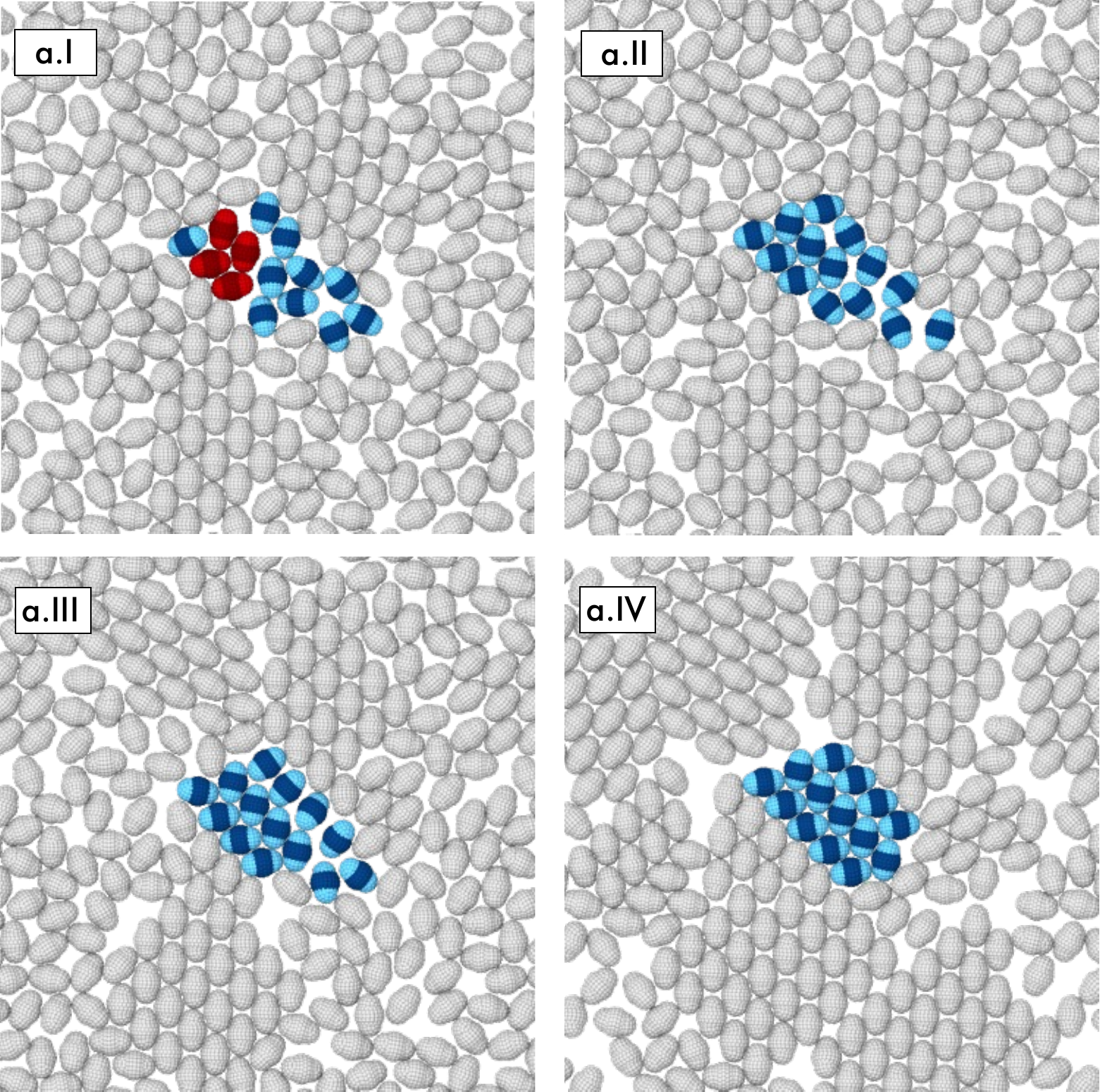} 
	 \caption{(a.I-IV) Representative time-lapse simulation snapshots showing the formation of a herringbone crystal in a system of triblock colloids with aspect ratio $\lambda=1.52$, depletant-to-colloid size ratio $q=0.05$, depletant packing fraction $\eta=1.05$ and colloid number density  $\rho=N_{c}\sigma^{2}/A=0.028$ for the largest crystal grain identified in the last timestep of the simulation. Red colloids highlight the seed that will be  part of the herringbone crystal in the last timestep of the simulation.
  }
	\label{fig:herringbone}
\end{figure*}

Finally, we also identify the seed for the herringbone phase, as shown in Fig.~\ref{fig:herringbone}.

\begin{figure*}[ht!]
\centering
	 \includegraphics [width=1\textwidth]{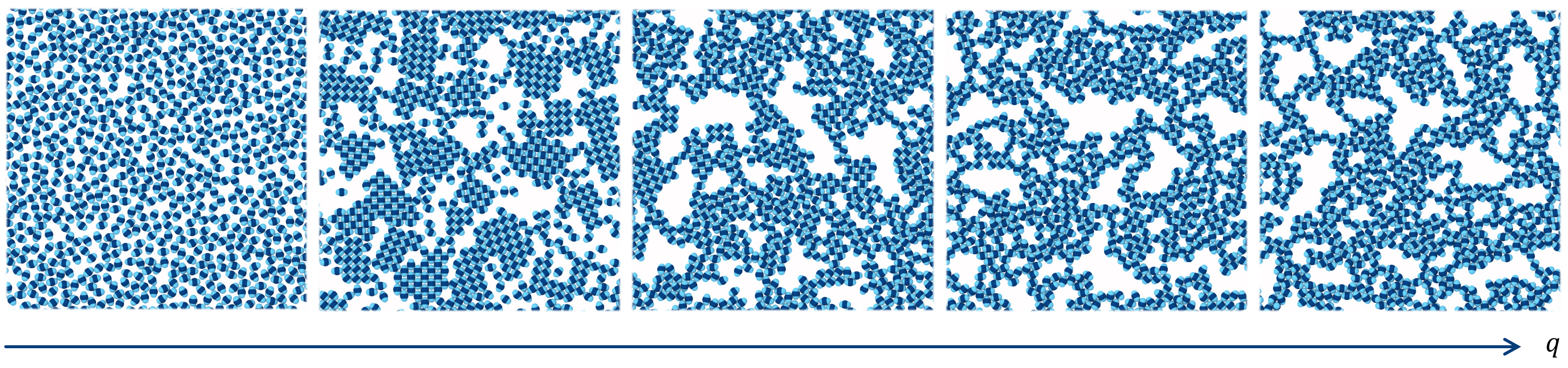} 
	 \caption{Representative simulation snapshots showing a system of triblock  colloids with aspect ratio $\lambda=1.52$ for increasing depletant-to-colloid size ratio $q=0.05, 0.075, 0.1, 0.125,$ and $0.15$ (from left to right) at a fixed depletant packing fraction $\eta=1.05$ and colloid number density $\rho=N_{c}\sigma^{2}/A=0.022$.}
	\label{fig:disordered}
\end{figure*} 

\section*{Disordered phases}

We report in Fig.~\ref{fig:disordered} representative simulation snapshots showing the different phases obtained for triblock  colloids with aspect ratio $\lambda=1.52$ for fixed depletant packing fraction $\eta$ and colloid number density $\rho=N_{c}\sigma^{2}/A$ and varying depletant-to-colloid sizes $q$. The colloids, at low $q$, retain a liquid-like behavior before forming crystal grains and  percolating network structures at higher $q$. At the highest $q$ value investigated, there is no crystalline order and the overall structure resembles that of a percolating gel.

\bibliography{main} 
\bibliographystyle{rsc} 

\end{document}